# On Brambles, Grid-Like Minors, and Parameterized Intractability of Monadic Second-Order Logic

Stephan Kreutzer
University of Oxford
kreutzer@comlab.ox.ac.uk

Siamak Tazari
Humboldt Universität zu Berlin
tazari@informatik.hu-berlin.de

**Abstract**

Brambles were introduced as the dual notion to treewidth, one of the most central concepts of the graph minor theory of Robertson and Seymour. Recently, Grohe and Marx showed that there are graphs $G$, in which every bramble of order larger than the square root of the treewidth is of exponential size in $|G|$. On the positive side, they show the existence of polynomial-sized brambles of the order of the square root of the treewidth, up to log factors. We provide the first polynomial time algorithm to construct a bramble in general graphs and achieve this bound, up to log-factors. We use this algorithm to construct grid-like minors, a replacement structure for grid-minors recently introduced by Reed and Wood, in polynomial time. Using the grid-like minors, we introduce the notion of a perfect bramble and an algorithm to find one in polynomial time. Perfect brambles are brambles with a particularly simple structure and they also provide us with a subgraph that has bounded degree and still large treewidth; we use them to obtain a meta-theorem on deciding certain parameterized subgraph-closed problems on general graphs in time singly exponential in the parameter; the only other result with a similar flavor that is known to us is due to Demaine and Hajiaghayi and obtains a doubly-exponential bound on the parameter (albeit, for a more general class of parameterized problems).

The second part of our work deals with providing a lower bound to Courcelle's famous theorem from almost two decades ago, stating that every graph property that can be expressed by a sentence in monadic second-order logic (MSO), can be decided by a linear time algorithm on classes of graphs of bounded treewidth. Whereas much work has been done on designing, improving, and applying algorithms on graphs of bounded treewidth, not much is known on the side of lower bounds: what bound on the treewidth of a class of graphs "forbids" polynomial-time parameterized algorithms to decide MSO-sentences? This question has only recently received attention with the first systematic study appearing in [Kreutzer 2009]. Using our results from the first part of our work we can improve on it significantly and establish a strong lower bound for Courcelle's theorem on classes of colored graphs.

# 1 Introduction

One of the deepest and most far-reaching theories of the recent 20 years in the realm of discrete mathematics and theoretical computer science is the *graph minor theory* of Robertson and Seymour. Over a course of over 20 papers, they prove the seminal graph minor theorem but perhaps even more importantly, develop a powerful and vast toolkit of concepts and ideas to handle graphs and understand their structure; indeed, a huge body of work has evolved that applies and extends these ideas in various fields of discrete mathematics and computer science. One of the most central concepts, introduced early on, is the notion of *treewidth*[1] [RS86b]. Treewidth has obtained immense attention ever since, especially because many NP-hard problems can be handled efficiently on graphs of bounded treewidth (e.g. all problems that can be defined in *monadic second-order logic* [Cou90]).

The dual notion to treewidth is the concept of a *bramble* [ST93, Ree97]; a bramble of large order is a witness for large treewidth. It turns out that so far, brambles have received far less attention than tree decompositions; perhaps the reason is that brambles can look quite complex and do not necessarily have a "nice" structure to be dealt with reasonably. Indeed, Robertson and Seymour figured out that there are certain brambles with "very nice" structure that are much more useful than general brambles: namely, a *grid-minor of large order*. In fact, Robertson and Seymour show that a graph has bounded treewidth if and only if it excludes a fixed grid as a minor [RS86a]. A grid is a canonical planar graph and the existence of large grids has various algorithmic and non-algorithmic applications and implications, e.g. [RS95, Epp00, Gro04, DFHT05, Gro07b, CSH08, Kre09]. However, the best known bounds relating treewidth and grid-minors are the following:

**Theorem 1.1.** *([RST94]) Every graph with treewidth at least $20^{2\ell^5}$ contains an $\ell \times \ell$-grid as a minor. There are graphs of treewidth $\ell^2 \log \ell$ that do not contain an $\ell \times \ell$-grid as a minor.*

So, there is a huge gap between the known lower and upper bounds of this theorem; Robertson and Seymour conjecture that the true value should be closer to the lower bound, i.e. that every graph should have a grid of order polynomial in the treewidth. Recently, Reed and Wood [RW08] attacked this problem by loosening the requirement for the bramble to be a grid; instead, they define a structure that they call a *grid-like minor*, as a replacement structure for a grid-minor, and prove that every graph does indeed contain a grid-like minor of order polynomial in the treewidth.

All of the results regarding brambles, grid-minors, and grid-like minors mentioned above are *existential*; to the best of our knowledge, it is not known so far how to *efficiently* construct *any* bramble of large order even when a tree decomposition of optimal width is given. It was not even studied up until recently, how large a bramble of the order of the treewidth can be; Grohe and Marx [GM09] showed that there exist brambles of size polynomial in the size of the graph whose order is roughly the square root of the treewidth (up to $\log$-factors); but they also show that there exist graphs, so that any bramble of order larger than the square root of the treewidth has size *exponential* in the size of the graph.

**Constructing Brambles.** We provide the first polynomial-time algorithm to compute a bramble that is guaranteed to have the order of the square-root of the treewidth, up to $\log$-factors, hence almost matching the best possible theoretical bound for polynomial-sized brambles. Our approach is based on the proof given in [GM09] but additionally, involves the approximation algorithms for treewidth, balanced separators, and sparse separators, which in turn are based on linear and semi-definite programming methods to obtain low-distortion metric embeddings of graphs [LR88, BGHK95, FHL08]. Even though we do not need to get into all of these topics in this work, it is interesting to note that it is a combination of all of these that finally gives rise to our algorithm. We also obtain an alternative (simpler) algorithm to construct a bramble of smaller size but lower order; in order to do so, we introduce the notion of a *k-web*, a structure that is similar to what Diestel et al. [DGJT99] denote by a *k-mesh*, and show that it can be computed by a polynomial time algorithm.

---
[1] see the next section for definitions.

Recently, Chapelle et al. [CMT09] presented an algorithm that computes a bramble of the order of the treewidth in time $\mathcal{O}(n^{k+4})$, where $n$ is the size of the graph and $k$ the treewidth; hence, they obtain brambles of optimal order but naturally, they need exponential time in order to do so. We would also like to mention a result by Bodlaender et al. [BGK05] that provide a polynomial-time *heuristic* to compute brambles in graphs; they use their algorithm for some computational experiments but do not prove any bounds on the order of the bramble they obtain.

**Constructing Grid-Like Minors.** Afterwards, we turn our attention to grid-like minors and present the first polynomial-time algorithm to construct a grid-like minor of large order in general graphs. Again, our method is based on the original existence proof of [RW08] but involves a number of new ideas and techniques, most notably the following: first, we make use of $k$-webs instead of brambles, and second, we (non-trivially) apply the very recent result of Moser [Mos09] that provides a certain algorithmic version of the Lovász Local Lemma. These two ideas make it possible that the algorithmic bound that we obtain (i.e. the order of the grid-like minor that we construct), is very close to the existential bound proved by Reed and Wood; if we would "just" use our bramble algorithm and proceed as in the original proof, the exponents would have about tripled. Also, we affirmatively answer a question by Reed and Wood [RW08] on whether the Local Lemma can be improved algorithmically for this application.

**Perfect Brambles.** As a first application of our results, we define the notion of a *perfect bramble* as a perhaps somewhat more "handy" replacement for grid-minors. Most notably, a perfect bramble defines a subgraph that has *bounded degree*, *large treewidth*, and has the property that every vertex appears in *at most* 2 bramble elements. We show that every graph contains a perfect bramble of order polynomial in the treewidth and that such a bramble can be computed in polynomial time. This shows that if the upper bound in Theorem 1.1 is to be improved to a polynomial, it is sufficient to prove it for perfect brambles.

**A Meta-Theorem.** Moreover, we present a *meta theorem* on perfect brambles: we show that essentially any graph parameter that is *subgraph monotone* and is *large on a perfect bramble*, can be decided in time $\mathcal{O}(2^{\text{poly}(k)} \text{poly}(n))$ and that a *witness* can be provided in the same time bound; here $n$ is the size of the input and $k$ is the size of the parameter. In the language of *parameterized complexity theory*, our result states that such parameters are *fixed-parameter tractable (fpt)* by a singly exponential fpt-algorithm.

One of the most important consequences of the graph minor theorem of Robertson and Seymour [RS95, RS04, FL88] is the following: for a given graph $G$ and parameter $\pi(G)$ that is *minor monotone*, one can decide if $\pi(G) \leq k$, in $\mathcal{O}(f(k)n^3)$-fpt time, where $f$ is an arbitrary function. This is, of course, a very general and very powerful theorem but there is a price to be paid: (i) for any such parameter, an algorithm is known to exist, but the algorithm itself can not be known in general; (ii) the theorem gives a *non-uniform* algorithm, meaning there is a different algorithm for every value of $k$; (iii) the function $f(k)$ is, in general, not computable and can be arbitrarily large. Frick and Grohe [FG01b] proved explicit bounds for certain graph classes and parameters that are definable in first-order logic, though the bounds were still non-elementary. Demaine and Hajiaghayi [DH07] proved a bound of $\mathcal{O}(2^{2^{\text{poly}(k)}} \text{poly}(n))$ for general graphs, when the considered parameter fulfills a few additional constraints. They use the grid-minor theorem for general graphs, together with ideas from the bidimensionality theory [DFHT05], to obtain this bound. By using a perfect bramble instead of a grid-minor, we can improve this bound to be singly-exponential in $k$, although the additional constraints that we require are somewhat stronger than the ones in [DH07]; still, our technique can be applied to many problems, for which their technique also applies.

**On Monadic Second Order Logic.** Another very well known meta-theorem, this time from logic, is Courcelle's famous result that every graph property definable in monadic second-order logic with quantifi-cation over sets of vertices and sets of edges (MSO$_2$) can be decided in linear time on any class of graphs of bounded treewidth [Cou90]. This immediately implies linear time algorithms for a wide range of problems from deciding whether a graph has a Hamiltonian cycle to 3-Colorability to parameterized algorithms for problems such as Dominating Set and most other covering problems. Following Courcelle's theorem, a range of other *algorithmic meta-theorems* have been obtained for more general classes of graphs, e.g. [FG01a, FG01b, DGKS06, DGK07]. See also recent surveys [Gro07a, Kre09] on the topic. More recently, the search for strong algorithmic meta-theorems based on logic has inspired work on parameterized graph algorithms, for instance in the work on meta-kernalization [BFL$^+$09].

Courcelle's theorem provides an easy way of proving that a problem can be solved efficiently on graph classes of bounded treewidth and has been used intensively in the literature. An obvious question is whether it is tight or can be extended to graph classes of unbounded treewidth, a natural choice being for instance the class $\mathcal{C}$ of graphs $G$ with treewidth $\mathrm{tw}(G) \leq \log |G|$. We say that the treewidth of $\mathcal{C}$ is bounded by $\log n$ or, more generally, by $\log^c n$ if $G \in \mathcal{C}$ implies $\mathrm{tw}(G) \leq \log^c n$, where $c$ is a constant.

The first systematic study of this question appears in [Kre09] where classes of graphs are studied whose treewidth is not bounded poly-logarithmically, or more precisely, not bounded by $\log^c n$, for some small constant $c$. The main result in [Kre09] essentially says that if $\mathcal{C}$ is a class of colored graphs whose treewidth is not bounded by $\log^c n$, then Courcelle's theorem does not extend to $\mathcal{C}$ (see Section 6 for details). However, [Kre09] only refers to classes which are called *constructible*, which essentially says that in graphs $G \in \mathcal{C}$ grid-like minors can be computed in polynomial time. The results of Section 4 remove this condition and establish a very strong lower bound for the complexity of monadic second-order logic. We show that, with respect to colored graphs, Courcelle's theorem is rather tight and can not be extended to classes of graphs of treewidth bounded by $\log^c n$ for $c > 24$.

**Organization**. We start by stating some preliminary notions and proceed with the above mentioned topics, in the given order.

## 2 Preliminaries

We usually denote graphs by letters $G, H$, and refer to their vertex/edge sets by $V(G)$ and $E(G)$, respectively. Unless otherwise mentioned, our graphs have $n$ vertices and $m$ edges. For a subset $U \subseteq V(G)$, we write $G[U]$ to denote the subgraph of $G$ induced by $U$. For an edge $e = uv$, we define the operation of *contracting* $e$ as identifying $u$ and $v$ and removing all loops and duplicate edges. A graph $H$ is a *minor* of $G$ if it can be obtained from $G$ by a series of vertex and edge deletions and contractions. A *model* of $H$ in $G$ is a map that assigns to every vertex of $H$, a connected subgraph of $G$, such that the images of the vertices of $H$ are all disjoint in $G$ and there is an edge between them if there is an edge between the corresponding vertices in $H$. A graph $H$ is a minor of $G$ if and only if $G$ contains a model of $H$. A *subdivision* of a graph $H$ is a graph that is obtained from $H$ by iteratively replacing some edges by paths of length 2. $H$ is a *topological minor* of $G$ if a subdivision of $H$ is a subgraph of $G$. A topological minor of $G$ is also a minor of $G$ but the reverse is not true in general. We refer the reader to [Die05] for more background on graph theory.

A *tree decomposition* of a graph $G$ is a pair $(T, \mathcal{B})$, where $T$ is a tree and $\mathcal{B} = \{B_i | i \in V(T)\}$ is a family of subsets of $V(G)$, called *bags*, such that (i) every vertex of $G$ appears in some bag of $\mathcal{B}$; (ii) for every edge $e = uv$ of $G$, there exists a bag that contains both $u$ and $v$; (iii) for every vertex $v$ of $G$, the set of bags that contain $v$ form a connected subtree $T_v$ of $T$. The *width* of a tree decomposition is the maximum size of a bag in $\mathcal{B}$ minus 1. The *treewidth* of a graph $G$, denoted by $\mathrm{tw}(G)$, is the minimum width over all possible tree decompositions of $G$. Let $f : \mathbb{N} \to \mathbb{N}$ be a function and $\mathcal{C}$ be a class of graphs. The treewidth of $\mathcal{C}$ is *bounded by* $f$, if $\mathrm{tw}(G) \leq f(|G|)$ for all $G \in \mathcal{C}$. $\mathcal{C}$ has *bounded treewidth* if its treewidth is bounded

by a constant.

**Definition 2.1.** *Let $G$ be a graph. Two subgraphs $B, B'$ of $G$ touch if they share a vertex or if there is an edge $e \in E(G)$ joining $B$ and $B'$. A bramble in $G$ is a set $\mathscr{B}$ of connected subgraphs of $V(G)$ such that any two $B, B' \in \mathscr{B}$ touch. The subgraphs in $\mathscr{B}$ are called bramble elements. A set $S \subseteq V(G)$ is a hitting set for $\mathscr{B}$ if it intersects every element of $\mathscr{B}$. The order of $\mathscr{B}$ is the minimum size of a hitting set.*

The canonical example of a bramble is the set of crosses (union of a row and a column) of an $\ell \times \ell$-grid. The following theorem shows the duality of treewidth and brambles:

**Theorem 2.2.** *([ST93]) A graph $G$ has treewidth at least $\ell$ if and only if $G$ contains a bramble of order at least $\ell + 1$.*

For the algorithmic purposes of this work, the following theorem due to Grohe and Marx is of high significance; it essentially says that if we are looking for a polynomial-sized bramble, the best order we can hope for is about the square-root of the treewidth:

**Theorem 2.3.** *([GM09])*
  (i) *Every $n$-vertex graph $G$ of treewidth $k$ has a bramble of order $\Omega(\frac{\sqrt{k}}{\log^2 k})$ and size $\mathcal{O}(k^{\frac{3}{2}} \cdot \ln n)$.*
  (ii) *There is a family $(G_k)_{k \geq 1}$ of graphs such that:*
     - *$|V(G_k)| = \mathcal{O}(k)$ and $E(G_k) = \mathcal{O}(k)$ for every $k \geq 1$;*
     - *$\operatorname{tw}(G_k) \geq k$ for every $k \geq 1$;*
     - *for every $\varepsilon > 0$ and $k \geq 1$, every bramble of $G_k$ of order at least $k^{\frac{1}{2}+\varepsilon}$ has size at least $2^{\Omega(k^\varepsilon)}$.*

We defer the definition of a *grid-like minor* to Section 4. Finally, we briefly review some basic notions of *parameterized complexity theory* [DF99, FG06]. We use the term $\operatorname{poly}(n)$ to denote some polynomial function in $n$ (often written as $n^{\mathcal{O}(1)}$ in the literature). A *parameter* for a problem is a function that assigns a natural number to every instance of the problem. Unless otherwise mentioned, we denote the problem size by $n$ and the parameter value by $k$. A problem is saeid to be *fixed-parameter tractable (fpt)*, if it can be solved by an algorithm in time $\mathcal{O}(f(k)\operatorname{poly}(n))$, for some computable function $f$. The class FPT is the set of all parameterized problems that are fixed-parameter tractable. The class XP is the set of all parameterized problems that can be solved by an algorithm in time $\mathcal{O}(n^{f(k)})$, for a computable function $f$. Clearly, FPT $\subseteq$ XP; Downey and Fellows [DF99] showed that, in fact, FPT $\neq$ XP. We say a parameterized problem can be solved by a *singly exponential* FPT algorithm if there is an algorithm for it with running time $\mathcal{O}(2^{\operatorname{poly}(k)} \operatorname{poly}(n))$.

## 3 Constructing Brambles and Webs

In this section, we show two different methods to construct a bramble in a graph. The first one is based on a randomized construction by Grohe and Marx [GM09]; it turns out that their proof of the existence of a large bramble can be made into a polynomial-time algorithm if one can find a large set whose sparsest cut is "not sparse". In order to find such a set, we use the ideas in the approximation algorithm for treewidth, where sparse cuts are used to construct balanced cuts and balanced cuts are, in turn, used to construct a tree decomposition. Our main idea is to make the approximation algorithm fail in a way that it provides us with the desired set.

In Sections 3.3–3.4, we introduce various notions of *k-webs* and show that they can be found in polynomial time. Our second bramble construction uses a $k$-web in order to obtain a bramble whose order is less than the order achieved by our first construction but whose size does not depend on $n$. It also has the advantage that it provides us with a deterministic and somewhat simpler algorithm to construct a bramble. We also need $k$-webs in Section 4 to construct grid-like minors.

We often need the approximation algorithm for treewidth, due to Bodlaender et al. [BGHK95] and its improved approximation ratio by Feige et al. [FHL08]. We summarize their result in the following lemma:

**Lemma 3.1.** *Given a graph $G$ of treewidth $k^\star$, there is a polynomial-time algorithm that constructs a tree decomposition of width $k_1$, such that for constants $c_0, c_1, c_2$, we have*

(i) $\frac{k_1}{c_0 \sqrt{\log k_1}} \leq k^\star \leq k_1 \leq c_0 k^\star \sqrt{\log k^\star}$;

(ii) *by setting* $k_2 = \left\lfloor \frac{k_1}{c_0 \sqrt{\log k_1}} \right\rfloor$, *we also obtain* $\frac{k^\star}{c_1 \sqrt{\log k^\star}} \leq k_2 \leq k^\star \leq c_2 k_2 \sqrt{\log k_2}$.

## 3.1 Finding A Large Set Lacking Sparse Separators

A *separator* of a graph $G$ is a partition of its vertices into three classes $(A, B, S)$, so that there are no edges between $A$ and $B$. The *size* of a separator is the size of the set $S$. For a subset $W \subseteq V(G)$, we say that a separator is $\gamma$-*balanced* or just a $\gamma$-*separator* with respect to $W$, if $|A \cap W|, |B \cap W| \leq \gamma |W|$. The treewidth of a graph is closely related to the existence of balanced separators:

**Lemma 3.2.** *(see e.g. [Ree97, FG06])*

(i) *If $G$ has treewidth greater than $3k$, then there is a set $W \subseteq V(G)$ of size exactly $2k + 1$ having no balanced $\frac{1}{2}$-separator of size $k$;*

(ii) *if $G$ has treewidth at most $k$, then every $W \subseteq V(G)$ has a balanced $\frac{1}{2}$-separator of size $k + 1$.*

The *sparsity* of a separator $(A, B, S)$ with respect to $W$ is defined as

$$\alpha^W(A, B, S) = \frac{|S|}{|(A \cup S) \cap W| \cdot |(B \cup S) \cap W|}.$$

We denote by $\alpha^W(G)$ the minimum of $\alpha^W(A, B, S)$ for every separator $(A, B, S)$. It is easy to see that for every connected $G$ and nonempty $W$, $\frac{1}{|W|^2} \leq \alpha^W(G) \leq \frac{1}{|W|}$. We are interested in a set $W$ with *no sparse separator*, i.e. where the sparsity of the sparsest cut is close to the maximum. Grohe and Marx [GM09] showed that the non-existence of balanced separators can guarantee the existence of such a set $W$:

**Lemma 3.3.** *([GM09]) If $|W| = 2k + 1$ and $W$ has no balanced separator of size $k$ in a graph $G$, then $\alpha^W(G) \geq \frac{1}{4k+1}$.*

The proof of Lemma 3.2 is algorithmic, but the algorithm is not polynomial-time since deciding if a (set in a) graph has a balanced separator of size $k$ is an NP-complete problem. Hence, we have to work with approximations. On the other hand, Grohe and Marx note that Lemma 3.3 does not remain true for larger $W$ by showing an example with $|W| = 4k$ and $\alpha^W(G) = \mathcal{O}(1/k^2)$; so, if we work with approximations, we can not use this lemma directly. We show in this section how to circumvent these problems by presenting a polynomial-time algorithm to find a large set $W$ with no sparse separator. Our algorithm follows the framework of approximating balanced separators by using sparse separators, as introduced by Leighton and Rao [LR88]. Additionally, we make use of the following two results:

**Lemma 3.4.** *(Feige et al. [FHL08]) Let $G$ be a connected graph, $W \subseteq V(G)$, and $T$ be the optimal $\frac{2}{3}$-separator of $W$ in $G$. There exists a polynomial-time algorithm that computes a separator $(A, B, S)$ of $G$, so that $\alpha^W(A, B, S) \leq \beta_0 \alpha^W(G) \sqrt{\log |T|}$, for some constant $\beta_0$.*

**Lemma 3.5.** *(adapted from Bodlaender et al. [BGHK95]) Let $G$ be a graph and $s \in \mathbb{N}$ be given. Suppose that for any connected subset $U$ of $V(G)$ and given set $W \subseteq U$ with $|W| = 4s$, there exists a $\frac{3}{4}$-separator of $W$ in $U$ of size at most $s$ and that such a separator can be found in polynomial time. Then the treewidth of $G$ is at most $5s$ and an according tree decomposition can be found in polynomial time.*

Now we can state our main technical lemma of this section; the proof is based on a technique from [LR88]:

**Lemma 3.6.** *Let $G$ be a graph of treewidth $k^\star$, $U_0$ a connected subset of $V(G)$ and $W_0 \subseteq U_0$ with $|W_0| = 4\beta_1 k$, where $\beta_1$ is a constant and $k$ a parameter. Then there exists a polynomial-time algorithm that either finds a $\frac{3}{4}$-separator of $W_0$ in $U_0$ of size at most $\beta_1 k$; or determines that $k < \frac{4}{3}k^\star\sqrt{\log k^\star}$ and returns a connected subset $U$ of $U_0$ and a subset $W \subseteq U$ with $|W| \geq 3\beta_1 k$, so that $\alpha^W(U) \geq \frac{1}{\beta_2 k^\star \log k^\star}$, where $\beta_2$ is a constant.*

*Proof.* We denote by $|X|_W$, the number of elements of $W$ in a set $X$. In our algorithm, we maintain a current component $U$ initialized to $U_0$, a current set $W \subseteq U$, $W \subseteq W_0$ initialized to $W_0$, and a current separator $S$ initialized to $\emptyset$. We keep the invariant that $|W| \geq \frac{3}{4}|W_0| = 3\beta_1 k$. In each iteration, we do the following: first, we find a separator $(A', B', S')$ of $W$ in $U$ as guaranteed by Lemma 3.4. Then, we know that $\alpha^W(A', B', S') \leq \beta_0 \alpha^W(U)\sqrt{\log |T|}$, where $(A_T, B_T, T)$ is the optimal $\frac{2}{3}$-separator of $W$ in $U$. Note that $T$ is at most the size of the optimal $\frac{1}{2}$-separator and hence, is at most $k^\star + 1$, by Lemma 3.2. Now, we have

$$\frac{|S'|}{|A' \cup S'|_W \cdot |B' \cup S'|_W} \leq \beta_0 \frac{|T|\sqrt{\log |T|}}{|A_T \cup T|_W \cdot |B_T \cup T|_W} \leq \beta_1 \frac{k^\star \sqrt{\log k^\star}}{|W|^2},$$

where the first inequality follows from the fact that $T$ is *some* separator of $W$ in $U$ and so, not sparser than the sparsest separator of $W$ in $U$; and the second inequality from $|A_T \cup T|_W, |B_T \cup T|_W \geq \frac{1}{3}|W|$ by requiring $\beta_1 \geq 18\beta_0$. It follows that $|S'| \leq \beta_1 k^\star \sqrt{\log k^\star} \frac{|B' \cup S'|_W}{|W|}$. We distinguish two cases:

**Case 1:** $|S'| > \beta_1 k \frac{|B' \cup S'|_W}{|W_0|}$. Then it must be that $k < \frac{4}{3}k^\star\sqrt{\log k^\star}$ and we have

$$\alpha^W(A', B', S') = \frac{|S'|}{|A' \cup S'|_W \cdot |B' \cup S'|_W} > \frac{\beta_1 k}{|A' \cup S'|_W \cdot |W_0|} \geq \frac{\beta_1 k}{|W_0|^2} = \frac{\beta_1 k}{16\beta_1^2 k^2} = \frac{1}{16\beta_1 k}$$

and hence,

$$\alpha^W(U) \geq \frac{\alpha^W(A', B', S')}{\beta_0 \sqrt{\log |T|}} \geq \frac{1}{22\beta_0\beta_1 k^\star \sqrt{\log k^\star}\sqrt{\log k^\star + 1}} \geq \frac{1}{\beta_2 k^\star \log k^\star},$$

for a constant $\beta_2 \geq 44\beta_0\beta_1$.

**Case 2:** $|S'| \leq \beta_1 k \frac{|B' \cup S'|_W}{|W_0|}$. We update our overall separator $S$ to be $S \cup S'$ and check if there exists a connected component $U'$ of $U \setminus S$ that still has more than a $\frac{3}{4}$-fraction of the elements of $W_0$. If so, we set $U = U'$ and $W = W_0 \cap U$ and repeat our algorithm. Otherwise $S$ is a $\frac{3}{4}$-separator of $W_0$ in $U_0$ and we claim that $|S| \leq \beta_1 k$: w.l.o.g we may always assume that $|A' \cup S'|_W \geq |B' \cup S'|_W$ and hence, after each iteration, the set $B' \cup S'$ is disgarded. So, the total sum, over all iterations, of the $|B' \cup S'|_W$ is at most $|W_0|$ and the claim follows. □

By setting $s = \beta_1 k$ in Lemma 3.5, we obtain a polynomial-time algorithm that given a graph $G$ and a parameter $k$, either finds a tree decomposition of $G$ of width at most $5\beta_1 k$ or returns sets $U$ and $W$ as specified in Lemma 3.6. Now, we can apply this algorithm with parameter $k = 2^i$ for $i = 0, 1, 2, \ldots$ to find the first $i$, so that it still fails in constructing a tree decomposition on $i$ but succeeds in doing so on $i + 1$. Hence, we have

**Lemma 3.7.** *There is a polynomial-time algorithm that given a graph G of treewidth $k^\star$, returns a number $k \in \mathbb{N}$, so that $\frac{k^\star}{10\beta_1} \leq k < \frac{4}{3}k^\star\sqrt{\log k^\star}$, together with a connected subset U of $V(G)$ and a set $W \subseteq U$ with $3\beta_1 k \leq |W| \leq 4\beta_1 k$, so that $\alpha^W(U) \geq \frac{1}{\beta_2 k^\star \log k^\star}$, where $\beta_1, \beta_2$ are constants.*

## 3.2 Randomized Construction of Brambles

Once we are able to find a large set with no sparse cuts in a graph, the rest of the probabilistic proof of Theorem 2.3 (i) in [GM09] becomes algorithmic. Given a set $W$ of vertices, a *concurrent vertex flow of value $\varepsilon$* is a collection of $|W|^2$ flows such that for every ordered pair $(u, v) \in W \times W$, there is a flow of value $\varepsilon$ between $u$ and $v$, and the total amount of flow going through each vertex is at most 1. A maximum concurrent vertex flow can be computed in polynomial time using linear programming techniques [FHL08].

The algorithm FIND-BRAMBLE is given below; steps (2)–(8) are reproduced from [GM09]. The basic ideas are as follows: first, we find a number $k$ and sets $U$ and $W_0$ as in Lemma 3.7; then we compute a maximum concurrent vertex flow on $W_0$; we select an arbitrary set $W \subseteq W_0$ of size $k$; afterwards, Grohe and Marx define a certain probability distribution on the paths between the vertices of $W$, based on the solution to the flow problem, and specify how to randomly pick and combine a number of these paths to construct, with high probability, a bramble $\mathcal{B}$.

**Algorithm** FIND-BRAMBLE($G$).
  *Input.*   an arbitrary graph $G$
  *Output.*  a bramble $\mathcal{B}$ in $G$

  1. apply Lemma 3.7 to obtain a number $k$, and sets $U, W_0 \subseteq V(G)$ as specified;
  2. compute a maximum concurrent vertex flow on $W_0$; let $p^{uv}$ denote the amount of flow that is sent from $u$ to $v$ along a path $p$;
  3. select $W \subseteq W_0$ with $|W| = k$ arbitrarily;
  4. let $d := \lfloor k^{3/2} \rfloor$ and $s := \lfloor \sqrt{k} \ln k \rfloor$; select sets $S_1, \ldots, S_d \subseteq W$, each of size $s$, uniformly and independently at random; let $S_i = \{u_{i,1}, \ldots, u_{i,s}\}$;
  5. for each $S_i$, select a vertex $z_i \in W \setminus S_i$ at random;
  6. for each $(u, v) \in W \times W$, let $\mathcal{P}_{uv}$ denote the set of all paths between $u$ and $v$; define a probability distribution on $\mathcal{P}_{uv}$ by setting the probability of $p \in \mathcal{P}_{uv}$ to be $\frac{p^{uv}}{\sum_{p' \in \mathcal{P}_{uv}} (p')^{uv}}$;
  7. for $i = 1, \ldots, s$ and $j = 1, \ldots, \lfloor \ln n \rfloor$ do
      - select one random path from each of $\mathcal{P}_{z_i, u_{i,1}}, \ldots, \mathcal{P}_{z_i, u_{i,s}}$ according to the probability distribution defined above; let $B_{i,j}$ be the union of these paths;
  8. return $\mathcal{B} := \bigcup_{i,j} B_{i,j}$.

Note that all the steps of the algorithm can be performed in polynomial time; in particular, the $p^{uv}$ are also variables in the linear programming formulation of the maximum concurrent flow problem and only a polynomial number of them will have nonzero value (cf. [FHL08]).

**Lemma 3.8.** *(adapted from Grohe and Marx [GM09]) With probability at least $1 - 1/k$, the set $\mathcal{B}$ constructed above is a bramble. With probability at least $1 - 1/n$, the order of this bramble is at least $\frac{k^{3/2} \alpha^{W_0}(U)}{\beta_3 \ln k \ln |W_0|}$, for a constant $\beta_3$.*

**Theorem 3.9.** *There exists a randomized polynomial time algorithm, that given a graph G of treewidth $k^\star$, constructs with high probability a bramble in G of size $\mathcal{O}(k^{\star 3/2} \ln k^\star \ln n)$ and order $\Omega(\frac{\sqrt{k^\star}}{\ln^3 k^\star})$.*

*Proof.* We apply the algorithm described in Section 3.2 and use Lemma 3.7 and Lemma 3.8 to bound the order and size of the bramble. For the size of the bramble, we know that $|\mathcal{B}| = \lfloor k^{3/2} \rfloor \lfloor \ln n \rfloor = O((k^\star \sqrt{\log k^\star})^{3/2} \ln n) = O(k^{\star 3/2} \ln k^\star \ln n)$. The order is at least

$$\geq \frac{k^{3/2} \cdot \alpha^{W_0}(U)}{\beta_3 \ln k \ln |W_0|} \geq \frac{k^{\star 3/2}}{\beta_4 k^\star \ln k^\star \ln^2 k} \geq \frac{\sqrt{k^\star}}{\beta_5 \ln^3 k^\star},$$

for appropriate constants $\beta_4, \beta_5 > 0$. □

Note that by a slight modification of the algorithm above, one can also construct a bramble of size $\mathcal{O}(k^{\star 3/2} \ln n)$ and order $\Omega(\frac{\sqrt{k^\star}}{\ln^4 k^\star})$.

## 3.3 Weak $k$-Webs

**Definition 3.10.** *A* weak $k$-web *of order $h$ in a graph $G$ is a set of $h$ disjoint trees $T_1, \ldots, T_h$, such that for all $1 \leq i < j \leq h$ there is a set $\mathcal{P}_{i,j}$ of $k$ disjoint paths connecting $T_i$ and $T_j$. If the trees $T_1, \ldots, T_h$ are all paths, we denote the resulting structure by a* weak $k$-web of paths *of order $h$.*

In [RW08], it is shown that any bramble of order at least $hk$, contains a weak $k$-web of paths of order $h$. They use this structure to show the existence of grid-like minors. Even though we provide a different proof for grid-like minors, we still include the following lemma as it might be of independent interest; also, note that one could use this lemma to construct grid-like minors, but it would result in worse bounds than what we obtain in Section 4.

**Lemma 3.11.** *There is a polynomial-time algorithm that given a bramble $\mathcal{B}$ of order at least $chk\sqrt{\log k}$ in a graph $G$, computes a weak $k$-web of paths of order $h$ in $G$, where $c$ is a constant.*

*Proof.* First, as in [RW08], we observe that one can find a simple path $P$ in $G$ that hits every element of $\mathcal{B}$ by a simple greedy algorithm: suppose by induction, that we have already constructed a path $P'$ that hits some elements of $\mathcal{B}$ and that there is one element $B \in \mathcal{B}$ that intersects $P'$ in only an endpoint $v$. If there is an element $B' \in \mathcal{B}$ that is not hit by $P'$, we extend $P'$ by a path $P_{vu} \subseteq B$, such that $P_{vu} \cap B' = \{u\}$; this is always possible, since $B$ and $B'$ touch. Furthermore $P_{vu}$ is otherwise disjoint from $P'$ and the extended path intersects $B'$ in only one vertex. Hence, our claim follows by induction.

Now, we move on $P$ from left to right and at each vertex $v$, we consider the sub-path $P_v$ and the sub-bramble $\mathcal{B}_v \subseteq \mathcal{B}$ that is hit by $P_v$. We can use the duality of brambles and tree decompositions and Lemma 3.1 to find a number $k_v$, such that $k_v \leq k_v^\star \leq c' k_v \sqrt{\log k_v}$, where $k_v^\star$ is the order of $\mathcal{B}_v$ and $c'$ is a constant. Now, let $uv$ be an edge of $P$, so that $k_u < k \leq k_v$. Note that $k_v^\star \leq k_u^\star + 1$ and hence, we obtain that the order of the sub-bramble $\mathcal{B}_v$ is at least $k$ and at most $ck\sqrt{\log k}$, for a properly defined constant $c$. We set $P_1 = P_v$ and $P' = P \setminus P_1$ and $\mathcal{B}' = \mathcal{B} \setminus \mathcal{B}_v$ and iterate this process on $P'$ and $\mathcal{B}'$. Since the order of the bramble $\mathcal{B}_v$, that is cut away in each iteration, is at most $ck\sqrt{logk}$ and since the order of $\mathcal{B}$ is at least $chk\sqrt{\log k}$, we indeed obtain at least $h$ disjoint paths $P_1, \ldots, P_h$ and brambles $\mathcal{B}_1, \ldots, \mathcal{B}_h$ each of order at least $k$, such that for all $i$, $P_i$ hits $\mathcal{B}_i$ and for $i < j$, $P_i$ does not hit $\mathcal{B}_j$. Reed and Wood [RW08] show that in this case, there exist at least $k$ disjoint paths between $P_i$ and $P_j$ for each $i < j$ and hence, the lemma is proven. □

**Corollary 3.12.** *For any $\varepsilon > 0$, there is a constant $c$, so that if for a graph $G$, we have $\mathrm{tw}(G) \geq ch^{2+\varepsilon}k^{2+\varepsilon}$, then $G$ contains a weak $k$-web of paths of order $h$ that can be constructed in randomized polynomial time.*

## 3.4 $k$-Webs

**Definition 3.13.** *A tree $T$ is* sub-cubic *if its maximum degree is at most $3$. A set $X \subseteq V(T)$ is called* flat *if every vertex $v \in X$ has degree at most $2$ in $T$.*

We will need the following lemma, whose simple proof is left for the reader.

**Lemma 3.14.** *Let $T$ be a sub-cubic tree and $X \subseteq V(T)$ be a set of $2 \cdot k \cdot l$ vertices, where $k, l \in \mathbb{N}$. Then there are $l$ disjoint sub-trees $T_1, \ldots, T_l$ of $T$ such that $|X \cap V(T_i)| = k$, for all $1 \leq i \leq l$.*

**Definition 3.15.** *A $k$-web of order $h$ in a graph $G$ is a collection $(T, (T_i)_{1 \leq i \leq h}, (A_i)_{1 \leq i \leq h}, B)$ of sub-graphs of $G$ such that*
  (i) *$T$ is a sub-cubic tree and $V(B \cap T) = \bigcup_{1 \leq i \leq h} V(A_i)$;*
  (ii) *$T_1, \ldots, T_h$ are disjoint subtrees of $T$ and for $1 \leq i \leq h$, $A_i \subseteq T_i$ is flat in $T$;*
  (iii) *for all $1 \leq i < j \leq h$ there is a set $\mathcal{P}_{i,j}$ of $k$ disjoint paths in $B$ connecting $A_i$ and $A_j$;*

Note that the main restriction of a $k$-web compared to a weak $k$-web is that the paths $\mathcal{P}_{i,j}$ are required to be disjoint from the trees $T_1, \ldots, T_h$ (except for their endpoints); on the other hand, the advantage of a weak $k$-web of paths is that all its trees are paths. Adapting a proof by Diestel et al. [DGJT99, Die05] we show that any graph of large enough treewidth contains a $k$-web of large order that can be computed in polynomial time.

**Lemma 3.16.** *Let $h, k \geq 1$ be integers. If $G$ has treewidth at least $(2 \cdot h + 1) \cdot k - 1$ then $G$ contains a $k$-web of order $h$. Furthermore, there is a polynomial time algorithm which, given $G, k, h$ either computes a tree decomposition of $G$ of width at most $(2 \cdot h + 1) \cdot k - 2$ or a $k$-web of order $h$ in $G$.*

*Proof.* W.l.o.g. we assume that $G$ is connected. Let $l := 2 \cdot k \cdot h$. A *pre-web* is a collection $\mathcal{W} := (U, \mathcal{D}, \{T_C : C \text{ is a component of } G - U\})$ where $U \subseteq V(G)$, $\mathcal{D} := (D, (B_t)_{t \in V(D)})$ is a tree decomposition of $G[U]$ of width at most $l + k - 2$ and for each component $C$ of $G \setminus U$, $T_C$ is a sub-cubic tree in $G \setminus C$ such that

  (i) there is a bag $B$ of $\mathcal{D}$ with $N(C) \subseteq B$;
  (ii) $N(C)$ is a flat subset of $V(T_C)$;
  (iii) $T$ has a leaf in $N(C)$ or $|T| = 1$ and $T \subseteq N(C)$.

$U$ is called the domain of the pre-web. The order of $\mathcal{W}$ is $|U|$. Inductively, we will construct a sequence of pre-webs of growing order until we either find a $k$-web of order $h$ or a pre-web with domain $V(G)$ and hence a tree decomposition of $G$ of width at most $l + k - 2$.

To initialize the algorithm choose a vertex $v \in V(G)$ and let $U := \{v\}$, $\mathcal{D} := (\{0\}, B_0 := \{v\})$ and $T_C := v$ for each component $C$ of $G - v$. Clearly, $(U, \mathcal{D}, \{T_C : C \text{ component of } G - v\})$ is a pre-web.

Suppose we have already constructed a pre-web $(U, \mathcal{D}, \{T_C : C \text{ component of } G \setminus U\})$. If $U = V(G)$ we are done. Otherwise, let $C$ be a component of $G \setminus U$ and let $T := T_C$. By assumption, there is a node $t \in V(D)$ with bag $B_t$, where $D$ is the tree underlying $\mathcal{D}$, such that $X := N(C) \subseteq B_t$.

If $|X| \leq l$ then let $v$ be a leaf of $T$ in $X$, which exists by assumption. Let $u \in V(C)$ be a neighbor of $v$ and set $U' := U \cup \{u\}$. Let $T' := T + \{u, v\}$ be the tree obtained from $T$ by adding $u$ as a new vertex joined to $v$. Further, let $\mathcal{D}'$ be the tree decomposition of $G[U']$ obtained from $\mathcal{D}$ by adding a new vertex $s$ with bag $B_s := X \cup \{u\}$ joined to $t$ in $\mathcal{D}'$. Now let $C'$ be a component of $G \setminus U'$. If $C' \cap C = \emptyset$ set $T'_{C'} := T_{C'}$. Otherwise, $C' \subseteq C$ and we set $T'_{C'}$ to be the minimal subtree of $T'$ containing $N(C')$. By construction, $N(C')$ contains $v$. Further, as $X = N(C)$ was flat in $T$, $N(C')$ is flat in $T'_{C'}$. Hence, $(U', \mathcal{D}', \{T'_{C'} : C' \text{ component of } G - U'\})$ is a pre-web of order $|U| + 1$.

Now suppose $|X| = l$. Let $T_1, \ldots, T_h$ be a collection of disjoint sub-trees of $T$ with $|V(T_i) \cap X| = k$, which exist by Lemma 3.14, and let $A_i := V(T_i) \cap X$. For each $1 \leq i < j \leq h$ compute a maximal set $\mathcal{P}_{i,j}$ of disjoint paths in $H := G[V(C) \cup A_i \cup A_j] \setminus E(G[A_i \cup A_j])$ joining $A_i$ and $A_j$. If all $P_{i,j}$ contain at least $k$ paths then $(T, (T_i)_{1 \leq i \leq h}, (A_i)_{1 \leq i \leq h}, C \cup N(C))$ is a $k$-web of order $h$ and we are done. Otherwise, let $A_i, A_j$ be such that $k' := |\mathcal{P}_{i,j}| < k$. By Menger's theorem, there is a set $S \subseteq V(H)$ of $k'$ vertices separating $A_i, A_j$ in $H$. Clearly, $S$ contains one vertex of each $P \in \mathcal{P}_{i,j}$. We denote by $P_s \in \mathcal{P}_{i,j}$ the path containing $s \in S$.

Let $X' := X \cup S$ and $U' := U \cup S$ and let $\mathcal{D}'$ be the tree decomposition of $G[U']$ obtained from $\mathcal{D}$ by adding a new vertex $r$ with bag $X'$ joined to $t$. By construction, $|X'| \leq |X| + |S| \leq l + k - 1$. Let $C'$ be a component of $G \setminus U$. If $C' \cap C = \emptyset$ set $T'_{C'} := T_{C'}$. Otherwise, $C' \subseteq C$ and $N(C') \subseteq X'$. Furthermore, $C'$ must have at least one neighbor $v$ in $S \cap C$ since $X$ does not separate $C'$ from $S \cap C$. By construction of $S$, $C'$ cannot have neighbors in both $A_i \setminus S$ and $A_j \setminus S$. W.l.o.g. we assume that $N(C') \cap A_i = \emptyset$. Let $T'_{C'}$ be the union of $T_C$ with all $A_i - S$-subpaths of $P_s$ for $s \in C \cap N(C')$. As these sub-paths start in $A_i \setminus S$ and have no inner vertices in $X'$, they do not meet $C'$. We claim that $\mathcal{W}' := (U', \mathcal{D}', \{T'_{C'} : C'$ component of $G \setminus U'\})$ is a pre-web. Clearly, $\mathcal{D}'$ is a tree decomposition of $G[U']$ of width at most $l + k - 2$. Furthermore, each tree $T'_{C'}$ is clearly sub-cubic. Now let $C'$ be a component of $G \setminus U'$. If $C' \cap C = \emptyset$, then $C'$ is also a component of $G \setminus U$ and hence $T'_{C'} = T_{C'}$ and therefore there is a bag $B_t$ in $\mathcal{D}$ with $N(C') \subseteq B_t$ and the additional conditions on $T_{C'}$ are met. Otherwise, $N(C') \subseteq X'$. Let $T := T'_{C'}$. Then $T$ contains a leaf in $X'$ (the vertex $v$ constructed above). The degree conditions imposed on $T$ are clearly met as well. Furthermore, $N(C')$ is a terminal subset of $T'_{C'}$. It follows that $\mathcal{W}'$ is a pre-web of order $|U'| > |U|$.

Obviously, the algorithm takes only a linear number of steps. Furthermore, each step can be computed in polynomial time. This concludes the proof. □

**Lemma 3.17.** *Let $k \geq 1$. If $G$ contains a $(k+1)$-web of order $k+1$ then the treewidth of $G$ is at least $k$.*

*Proof.* Let $(T, (T_i)_{1 \leq i \leq k+1}, (A_i)_{1 \leq i \leq k+1}, Z)$ be a $(k+1)$-web of order $k+1$ in $G$. Towards a contradiction, assume $G$ has a tree decomposition $(D, (B_t)_{t \in V(D)})$ of width $< k$. For an edge $st \in E(D)$, we denote by $D_{s-t}$ the subtree of $D - st$ that contains $s$ and by $B(D_{s-t})$, the union of the bags of $D_{s-t}$. We orient the edges of $D$ as follows. If $st \in E(D)$, let $I_s := \{T_i : T_i \subseteq B(D_{s-t})\}$ and define $I_t$ analogously; we orient the edge towards $s$ if $|I_s| \geq |I_t|$ and otherwise orient the edge towards $t$. As $D$ is acyclic, there must be a node $s^\star \in V(D)$ such that all incident edges are oriented towards $s^\star$. Now, for each edge $s^\star t \in E(D)$, $B(D_{s^\star - t})$ contains at least one $T_i$ completely; on the other hand, as $|B_{s^\star}| \leq k$, $B_{s^\star}$ can not contain a vertex of every $T_i$ and there must be an edge $s^\star t^\star \in E(D)$, so that $B(D_{t^\star - s^\star})$ also contains some $T_i$ completely. Let $T_i \subseteq B(D_{s^\star - t^\star})$ and $T_j \subseteq B(D_{t^\star - s^\star})$; but then, there are $k+1$ disjoint paths between $T_i$ and $T_j$ and each of these must have an inner vertex in $B_{s^\star} \cap B_{t^\star}$, which is impossible. □

**Corollary 3.18.** *There is a polynomial time algorithm which, given a graph $G$ either computes a $(k+1)$-web of order $k+1$ and thereby proves that $\mathrm{tw}(G) \geq k$ or a tree decomposition of $G$ of width $\mathcal{O}(k^2)$.*

### 3.5 Constructing a Bramble from a $k$-Web

In this subsection, we briefly sketch an alternative bramble construction that differs from the one in Section 3.2 in that its *size* does not involve $n$ but instead, its *order* is less[2].

**Lemma 3.19.** *Given a $k^2$-web of order $k$, one can construct a bramble of size $k^3$ and order $k$.*

*Proof.* Let $(T, (T_i)_{1 \leq i \leq k}, (A_i)_{1 \leq i \leq k}, B)$ be a $k^2$-web of order $k$ and let $\mathcal{P}_{i,j} = \{P^1_{ij}, \ldots, P^{k^2}_{ij}\}$ be the $k^2$ disjoint paths between $A_i$ and $A_j$. Let $\widehat{P^t_{ij}}$ be the path $P^t_{ij}$ without the last edge that connects it to $A_j$. Define

---
[2] The existence of such a bramble is briefly mentioned in [GM09] but it is not presented; thanks to Dániel Marx for a helpful discussion on this matter.

$B_i^t = T_i \cup \bigcup_{j=1}^{k} \widehat{P_{ij}^t}$, for $1 \leq i \leq k$ and $1 \leq t \leq k^2$, and let $\mathscr{B} = \bigcup_{i,t} B_i^t$. Then $\mathscr{B}$ is clearly a bramble of size $k^3$. Suppose there is a hitting set of $\mathscr{B}$ of order less than $k$; then there is an $i$, such that $T_i$ is not covered. Hence, for $1 \leq t \leq k^2$, $B_i^t$ must be covered using vertices in $\bigcup_{t,j} \widehat{P_{ij}^t}$; but note that any vertex in this union has degree at most $k$ and so, at least $k$ vertices are needed to cover all these $k^2$ sets. □

**Theorem 3.20.** *There exists a polynomial time algorithm that, given a graph $G$ of treewidth $k^\star$, constructs a bramble in $G$ of size $O(k^\star)$ and order $\Omega((\frac{k^\star}{\sqrt{\log k^\star}})^{1/3})$.*

*Proof.* By Lemma 3.1, we can compute $\frac{k^\star}{c_1\sqrt{\log k^\star}} \leq k_2 \leq k^\star$. We set $k = \frac{k_2^{1/3}}{2}$ and use Lemma 3.16 to obtain a $k^2$-web of order $k$ in $G$. Our claim then follows by Lemma 3.19. □

# 4 Constructing Grid-Like Minors

Let $\mathcal{P}$ and $\mathcal{Q}$ each be a set of disjoint connected subgraphs of a graph $G$. We denote by $\mathcal{I}(\mathcal{P}, \mathcal{Q})$ the *intersection graph* of $\mathcal{P}$ and $\mathcal{Q}$ defined as follows: $\mathcal{I}(\mathcal{P}, \mathcal{Q})$ is the bipartite graph that has one vertex for each element of $\mathcal{P}$ and $\mathcal{Q}$ and an edge between two vertices if the corresponding subgraphs intersect.

**Definition 4.1.** *Let $\mathcal{P}$ and $\mathcal{Q}$ be each a set of disjoint paths in a graph $G$. $\mathcal{P} \cup \mathcal{Q}$ is called a* grid-like minor *of order $\ell$ in $G$ if $\mathcal{I}(\mathcal{P}, \mathcal{Q})$ contains the complete graph $K_\ell$ as a minor. If the $K_\ell$-minor is, in fact, a topological minor, we call the structure a* topological grid-like minor *of order $\ell$.*

**Theorem 4.2.** *(Reed and Wood [RW08]) Every graph with treewidth at least $c\ell^4\sqrt{\log \ell}$ contains a grid-like minor of order $\ell$, for some constant $c$. Conversely, every graph that contains a grid-like minor of order $\ell$ has treewidth at least $\lceil \frac{\ell}{2} \rceil - 1$.*

The proof given in [RW08] is existential and proceeds as follows: first, using a large bramble, a weak $k$-web of paths is constructed; then for each pair of sets of disjoint paths in the $k$-web, it is checked whether their union contains a grid-like minor of large order; if this is not true for any pair, one can obtain a grid-like minor using the Lovász Local Lemma. In this section, we make their proof algorithmic by showing how the individual major steps of the proof can be performed in polynomial time. We show

**Theorem 4.3.** *There are constants $c_1, c_2, c_3, c_1', c_2'$, so that if a graph $G$ has*

(i) $\operatorname{tw}(G) \geq c_1 \ell^5$, *then $G$ contains either $K_\ell$ as a minor or a topological grid-like minor of order $\ell$;*

(ii) $\operatorname{tw}(G) \geq c_2 \ell^8$, *$G$ contains either $K_{\ell^2}$ as a minor or a $c_3 \ell^6$-web of order $4$ that contains a topological grid-like minor of order $\ell$;*

(iii) $\operatorname{tw}(G) \geq c_2 \ell^8$, *$G$ contains a topological grid-like minor of order $\ell$.*

*Furthermore, the corresponding objects can be constructed by a randomized algorithm with expected polynomial running time. If the bounds on the treewidth are loosened to $c_1' \ell^7$ and $c_2' \ell^{12}$, respectively, then a deterministic algorithm can be used.*

The first step of the proof in [RW08] is to find a weak $k$-web of paths; instead, we make use of a $k$-web as described in Section 3.4. We procede with the second main step of the algorithm.

## 4.1 Finding Complete Topological Minors

Once we have a $k$-web, we need to determine if the intersection graph of any pair of the disjoint paths contains a large complete graph as a minor. Thomason [Tho01] showed that if the average degree of a graph is at least $cp\sqrt{\log p}$, then the graph contains $K_p$ as a minor (and that this bound is tight). His proof is very complicated and it is not clear if it can be turned into a polynomial-time algorithm to actually find such a minor. However, if we are looking for a *topological minor*, we need an average degree of at least $cp^2$ and Bollobás and Thomason [BT98] show that this bound actually suffices. Furthermore, it turns out that their proof is, in fact, algorithmic:

**Theorem 4.4.** *(adapted from Bollobás and Thomason [BT98]) If a graph $G$ has average degree at least $cp^2$, for a constant $c$, then $G$ contains $K_p$ as a topological minor. Furthermore, a model of $K_p$ can be found in $G$ in polynomial time.*

Note that by the defition of a grid-like minor, we do not necessarily need a topological minor but we use them for two reasons: first, we know we can compute them in polynomial time; second, we need to have a topological minor in Section 6. The algorithm for Theorem 4.4 is given by Algorithm TOP-MINOR below. We refer for the full proof of correctness to the original paper [BT98] and just argue briefly that each of the steps can be performed in polynomial time.

**Algorithm** TOP-MINOR$(G, p)$.
  *Input.*    a graph $G$ with $e(G) \geq 256p^2 n$
  *Output.*   a topological minor $K_p$ in $G$
(in the following, the index $i$ ranges appropriately)
  1. find a subgraph $G_1$ of $G$ that is at least $128p^2$-connected;
  2. select an arbitrary set $X = \{x_1, \ldots, x_{3p}\}$ in $G_1$ and let $G_2 = G_1 \setminus X$;
  3. select $3p$ arbitrary disjoint sets $Y_1, \ldots, Y_{3p}$ in $G_2$ each of size $5p$, s.t. $Y_i$ consists of neighbours of $x_i$;
  4. find a set $Z \subseteq \bigcup Y_i$ of size $7p^2$ which is linkable;
  5. let $Z_i = Z \cap Y_i$ and select indices $j_1, \ldots, j_p$, so that $|Z_{j_i}| \geq p - 1$;
  6. return $\{x_{j_1}, \ldots, x_{j_p}\}$ together with the disjoint paths that exist between the $Z_{j_i}$.

The first step of the algorithm is due to a theorem of Mader [Mad72] (see also [Die05], Theorem 1.4.3) and can be computed as follows: we select $G_1$ as a minimal subgraph $G$, such that $n(G_1) \geq 256p^2$ and $e(G_1) \geq 256p^2(e(G_1) - 128p^2)$; we can start by setting $G_1 = G$ and deleting vertices and edges and finding minimum cuts to reduce $G_1$ as long as the desired properties are still satisfied. Clearly, these operations can all be performed in polynomial time and Mader shows that in the end, $G_1$ will be $128p^2$-connected.

The only major difficult step of the algorithm, is the 4th step. We call a set of vertices *linkable* if for any pairing of its elements, there exist disjoint paths between the given pairs. A graph is said to be $(k, \ell)$-*linked* if every set of $k$ vertices contains a subset of size $\ell$ which is linkable. Bollobás and Thomason show that $G_2$ is $(15p^2, 7p^2)$-linked and hence, that the set $Z$ exists. They proceed by first finding a minor $H$ of $G_2$ that has large minimum degree; this can be achieved by starting with $G_2$ and considering certain minimal minors (and minors thereof), all of which can be constructed in polynomial time by a series of edge deletions and contractions. By using this minor together with Menger's theorem, they are able to find certain disjoint paths and modify them until the desired properties are achieved. Since the application of Menger's theorem amounts to a maximum-flow computation, all of the steps can indeed be performed in polynomial time.

## 4.2 Algorithmic Application of the Lovász Local Lemma

Recall that a graph $G$ is called $d$-*degenerate* if every subgraph of $G$ has a vertex of degree at most $d$ and note that Theorem 4.4 implies that if $G$ does not contain $K_p$ as a topological minor, then $G$ is $cp^2$-degenerate, for a constant $c$. In this section, we prove the following lemma:

**Lemma 4.5.** *For some $r \geq 2$, let $V_1, \ldots, V_r$ be the color classes in an $r$-coloring of a graph $H$. Suppose that $2^{t+1} > |V_i| \geq 64(2r - 3)d \geq n := 2^t$ for all $1 \leq i \leq r$ and some integer $t$, and assume $H[V_i \cup V_j]$* is $d$-degenerate for $1 \leq i < j \leq r$. Then there exists a randomized algorithm that finds an independent set $\{x_1, \ldots, x_r\}$ of $H$, such that each $x_i \in V_i$, in expected time polynomial in $n$. Furthermore, if, instead, we have $n \geq r(r-1)d + 1$, then a deterministic algorithm can be used.

Reed and Wood [RW08] prove an *existential* version of this lemma, using the Lovász Local Lemma (LLL) [EL75] (with the slightly stronger bound of requiring $|V_i| \geq 2e(2r-3)d$, where $e$ is the base of the natural logarithm). They note that if $n \geq r(r-1)d + 1$, a simple minimum-degree greedy algorithm will work, and pose as an open question if this algorithmic bound can be improved. Our lemma above *answers this question affirmatively*. The proof is based on the following very recent algorithmic version of the LLL due to Moser [Mos09]; recall that a *t-CNF formula* is a boolean formula in conjunctive normal form where each clause has *exactly* $t$ literals:

**Theorem 4.6.** *([Mos09]) Let $F$ be a $t$-CNF formula such that each clause $C \in F$ has common variables with at most $2^{t-5} - 1$ other clauses. Then $F$ is satisfiable and there exists a randomized algorithm that finds a satisfying assignment to $F$ in expected time polynomial in $|F|$.*

Our proof of Lemma 4.5 is based on the idea of using for each set $V_i$, $t$ binary variables to encode the index of the vertex that is to be included in the independent set from this color class. This way, the forbidden pairs of selections can be expressed using exactly $2t$ variables, so that Theorem 4.6 can be applied.

*Proof of Lemma 4.5.* If for any $i$, we have $|V_i| > n$, we delete some vertices out of $V_i$, so as to have $|V_i| = n$, for all $i$; let $V_i = \{v_0^i, \ldots, v_{n-1}^i\}$. Note that deleting vertices does not change the degeneracy assumption. We construct a $2t$-CNF formula $F$ as follows: we introduce variables $b_j^i$, for $1 \leq i \leq r$ and $0 \leq j < t$. We think of each sequence $b_{t-1}^i \ldots b_0^i$ as encoding an index $b^i$ in binary, so that $x_i = v_{b^i}^i$ is to be included in the independent set. For each edge $e = v_y^i v_z^j$, we add a clause $C_e$ to $F$ as follows: let $y_{t-1} \ldots y_0$ and $z_{t-1} \ldots z_0$ be the binary representations of $y$ and $z$ respectively. If $y_l$ is 0, we include the term $b_l^i$ in $C_e$, otherwise we include $\overline{b_l^i}$ in $C_e$ and act accordingly for $z$. This way, it is ensured that $v_y^i$ and $v_z^j$ are not selected simultaneously and we obtain that $C_e$ has size exactly $2t$.

Now the clause $C_e$ has common variables exactly with those clauses that are built by edges that have an endpoint in $V_i$ or $V_j$. There are at most $(2r-3) \cdot 2dn$ such edges; hence, the number of clauses that have a common variable with $C_e$, including $C_e$, can be bounded by $2(2r-3)d \cdot n < 2^{t-5} \cdot 2^t = 2^{2t-5}$. Thus, our claim follows by Theorem 4.6. As for a deterministic algorithm, recall that if $n$ is large enough, a simple minimum-degree greedy algorithm can be used. □

## 4.3 Putting Things Together

Starting with a (weak) $k$-web of order $h$, we consider the disjoint paths $\mathcal{P}_{i,j}$ between the pairs of trees from the web; note that these paths can be found by a simple max-flow computation in polynomial time. For each pair of these paths, we check if the average degree of the intersection graph is large; if so, we find a topological grid-like minor by Theorem 4.4; otherwise, we consider the intersection graph $\mathcal{I}$ of *all* the $r := \binom{h}{2}$ sets of paths; i.e. $\mathcal{I}$ is an $r$-partite graph, having a vertex for each path out of $\mathcal{P}_{i,j}$, for $1 \leq i < j \leq h$, and an edge between two vertices if the corresponding paths intersect. Now we can invoke Lemma 4.5 with $\mathcal{I}, r$ and $d := c_1 p^2$. We obtain

**Lemma 4.7.** *Let $G$ be a graph and let $T_1, \ldots, T_h$ be given to be the disjoint trees of a (weak) $k$-web of order $h$ in $G$ with $k \geq ch^2 p^2$, for a constant $c$. Then there exists a randomized algorithm with polynomial expected running time that finds, in $G$, either a topological grid-like minor of order $p$ or a set of $\binom{h}{2}$ disjoint paths $Q_{ij}, 1 \leq i < j \leq h$, so that $Q_{ij}$ connects $T_i$ to $T_j$. If $k \geq c'h^4p^2$, a deterministic algorithm also exists.*

By using the $k$-web of order $h$ that is guaranteed by Lemma 3.16 and setting $k = ch^2p^2$, we immediately obtain a randomized algorithm that given a graph $G$ of treewidth at least $ch^3p^2$ computes in $G$ either a model of $K_h$ or a topological grid-like minor of order $p$; a deterministic variant is obtained if $\text{tw}(G) \geq c'h^5p^2$. This observation, in turn, easily proves Theorem 4.3; we only sketch briefly, how claim (iii) is obtained from claim (ii): consider a graph $H$ that consists of $\ell$ "horizontal" paths and $\binom{\ell}{2}$ "vertical" edges, one connecting each pair of the horizontal paths. Then $H$ has less than $\ell^2$ vertices, has maximum degree 3, and any subdivision of $H$ is a topological grid-like minor of order $\ell$; now, any graph that has $K_{\ell^2}$ as a minor, has $H$ as a topological minor and hence, contains a topological grid-like minor of order $\ell$ (recall that if a graph $H$ has maximum degree 3 and is a minor of a graph $G$, then it is also a topological minor of $G$).

Note that by using the weak $k$-web of paths that as given by Corollary 3.12, one can also directly obtain a topological grid-like minor of order $h$ but the bounds would be worse than those obtained by Theorem 4.3.

## 5 Perfect Brambles and a Meta-Theorem

In this section, we define perfect brambles and show that certain parameterized problems can be decided efficiently using this notion as a replacement for grid-minors.

### 5.1 Perfect Brambles

**Definition 5.1.** *A bramble $\mathscr{B}$ in a graph $G$ is called* perfect *if*
  1. *any two $B, B' \in \mathscr{B}$ intersect;*
  2. *for every $v \in V(G)$ there are at most two elements of $\mathscr{B}$ that contain $v$;*
  3. *every vertex has degree at most 4 in $\bigcup \mathscr{B}$.*

Perfect brambles have some interesting properties, such as the ones given below.

**Lemma 5.2.** *Let $\mathscr{B} = \{B_1, \ldots, B_k\}$ be a perfect bramble and let $H = \bigcup \mathscr{B}$. Then we have*
  (i) *every element $B \in \mathscr{B}$ has at least $k - 1$ vertices;*
  (ii) *every element $B \in \mathscr{B}$ has at least $k - 2$ edges that do not appear in any other element of $\mathscr{B}$;*
  (iii) *$H$ has at least $\frac{k(k-1)}{2}$ vertices and at least $k(k - 2)$ edges;*
  (iv) *the order of $\mathscr{B}$ is* exactly $\lceil \frac{k}{2} \rceil$ *and hence, can be computed in linear time;*
  (v) *the treewidth of $H$ is at least $\lceil \frac{k}{2} \rceil - 1$.*

*Proof.* Claim (i), (ii), and (iii) follow from the fact that $B$ intersects $k - 1$ other elements of $\mathscr{B}$ and because of Property (ii) in Definition 5.1, an extra vertex is needed for each; also, at least $k - 2$ edges are needed to connect these at least $k - 1$ parts of $\mathscr{B}$ together. Since each vertex covers at most two elements of $\mathscr{B}$, at least $\frac{k}{2}$ vertices are needed for a complete hitting set; on the other hand, since each two elements of $\mathscr{B}$ meet at a vertex, $\frac{k}{2}$ vertices are also sufficient. This proves claims (iv) and (v). □

**Theorem 5.3.** *There are constants $c_1, c_2, c_3$, such that for any graph $G$, we have*
  (i) *if $\text{tw}(G) \geq c_1 k^4 \sqrt{\log k}$, then $G$ contains a perfect bramble of order $k$;*
  (ii) *if $\text{tw}(G) \geq c_2 k^5$, there is randomized algorithm with expected polynomial running time that finds a perfect bramble of order $k$ in $G$;*
  (iii) *if $\text{tw}(G) \geq c_3 k^7$, a deterministic algorithm for the same purpose exists.*

*Proof.* Consider a grid-like minor of order $2k$ in $G$; let $\mathcal{P}, \mathcal{Q}$ be the sets of disjoint paths, so that $\mathcal{I} = \mathcal{I}(\mathcal{P}, \mathcal{Q})$ contains $K_{2k}$ as a minor. Let $I_1, \ldots, I_{2k}$ be the connected subgraphs of $\mathcal{I}$ that define a model of $K_{2k}$. For each of these subgraphs $I_j$, we define a subgraph $B_j$ of $G$ that consists of the set of paths out of $\mathcal{P}$ and $\mathcal{Q}$ that are contained in $I_j$. Then $\mathscr{B} = \{B_1, \ldots, B_{2k}\}$ is a perfect bramble of order $k$; this can be checked straightforwardly by noting that (i) $\mathcal{P}$ and $\mathcal{Q}$ are each a set of disjoint paths; (ii) the sets $I_1, \ldots, I_{2k}$ are disjoint in $\mathcal{I}$ and there is an edge between any two of them; (iii) when there is an edge between two sets $I_i$ and $I_j$, it means that there is a path in $B_i$ and a path in $B_j$, one from $\mathcal{P}$ and one from $\mathcal{Q}$, such that these two intersect[3].

Also, consider a $K_{2k}$-minor as guaranteed by Theorem 4.3 (i) and constructed by Lemma 4.7. It consists of a number of subcubic trees $T_1, \ldots, T_{2k}$ and a number of dijoint paths $Q_{ij}, 1 \leq i < j \leq 2k$. For $1 \leq i \leq 2k$, we define a set $B_i$ to be the union of $T_i$ with "the first half" of each of the paths $Q_{ij}, 1 \leq j \leq k, j \neq i$, where "the first half" is defined as follows: for each path $Q_{ij}$, we select an arbitrary vertex $v_{ij}$ on $Q_{ij}$; the first half of a path $Q_{ij}$, starting at the tree $T_i$, is then the part of the path up to and including $v_{ij}$. Then, one can easily check that $\mathscr{B} = \{B_1, \ldots, B_{2k}\}$ is a perfect bramble of order $k$.

Now our claim follows by Theorems 4.2 and 4.3. □

**Corollary 5.4.** *For any graph $G$ of treewidth $k$, there exists a subgraph $H$ of $G$ with treewidth polynomial in $k$ and maximum degree $4$. Furthermore, $H$ can be computed in polynomial time.*

An interesting consequence of this corollary is that if the relation between treewidth and grid-minors is indeed polynomial (see Theorem 1.1), then it suffices to prove it only for graphs of bounded degree, in fact, only for perfect brambles.

## 5.2 A Meta-Theorem on Perfect Brambles

Let $\mathscr{G}$ denote the set of all graphs; we have the following theorem:

**Theorem 5.5.** *Let $c, \alpha > 0$ be constants, $G$ be a graph, and $\pi : \mathscr{G} \to \mathbb{N}$ be a parameter, such that*

*(i) if $H$ is a subgraph of $G$, then $\pi(H) \leq \pi(G)$;*
*(ii) on any graph $H = \bigcup \mathscr{B}$, where $\mathscr{B}$ is a perfect bramble of order $\ell$, $\pi(H) \geq c\ell^\alpha$;*
*(iii) given a tree decomposition of width $\ell$ on a graph $H$, $\pi(H)$ can be computed in time $\mathcal{O}(2^{\text{poly}(\ell)} \text{poly}(n))$;*

*then there is an algorithm with running time $\mathcal{O}(2^{\text{poly}(k)} \text{poly}(n))$ that decides if $\pi(G) \leq k$. Furthermore, if in (i), (ii), and (iii) above, a corresponding witness can be constructed in time $\mathcal{O}(2^{\text{poly}(k)} \text{poly}(n))$, then a witness, proving or disproving $\pi(G) \leq k$, can also be constructed in the given time.*

The idea of the proof is as follows: if the treewidth of $G$ is large enough, then $G$ contains a subgraph $H := \bigcup \mathscr{B}$, where $\mathscr{B}$ is a perfect bramble of large order, and hence, by conditions (i) and (ii), $\pi(G) \geq \pi(H) \geq k$; otherwise, the treewidth of $G$ is bounded by $\text{poly}(k)$; a tree decomposition can be computed using, say, the approximation algorithm of treewidth [BGHK95, FHL08] (see Lemma 3.1) or the fpt algorithm by Bodlaender [Bod96] (see also [FG06]), and a solution can be directly computed by condition (iii) of the Theorem. Using Lemma 5.2 one can see that our meta-theorem above can be applied to a variety of problems, such as vertex cover, edge dominating set (= minimum maximal matching), feedback vertex set, longest path, and maximum-leaf spanning tree. Whereas there already exist better fpt algorithms for these problems, we do not know of a unifying argument like in Theorem 5.5 that provides singly-exponential fpt algorithms for all these problems; also, this technique might be applicable to other problems, for which singly-exponential fpt algorithms are not known yet. But the main significance of the theorem resides in the reasons discussed in the introduction of this work, regarding the algorithmic application of the graph minor theorem. Also, the algorithmic nature of Theorem 5.3 makes it possible to actually construct a *witness*, as specified by Theorem 5.5; this was, in general, not achieved by previous results.

---

[3]A similar proof is also given in [RW08].

# 6 Parameterized Intractability of MSO$_2$ Model Checking

In this section we use the results established above to significantly improve on a lower bound on Courcelle's theorem for classes of coloured graphs proved in [Kre09]. We first need some notation. Throughout this section we will work with coloured graphs. Let $\Sigma := \{B_1, \ldots, B_k, C_1, \ldots, C_l\}$ be a set of colours, where the $B_i$ are colours of edges and the $C_i$ are colours of vertices. A $\Sigma$-*coloured graph*, or simply $\Sigma$-*graph*, is an undirected graph $G$ where every edge can be coloured by colours from $B_1, \ldots, B_k$ and every vertex can be coloured by colours from $C_1, \ldots, C_k$. In particular, we do not require any additional conditions such as edges having endpoints coloured in different ways. A class $\mathcal{C}$ of $\Sigma$-graphs is said to be closed under $\Sigma$-colourings if whenever $G \in \mathcal{C}$ and $G'$ is obtained from $G$ by recolouring, i.e. the underlying un-coloured graphs are isomorphic, then $G' \in \mathcal{C}$.

The class of formulas of *monadic second-order logic with edge set quantification* on $\Sigma$-coloured graphs, denoted MSO$_2[\Sigma]$, is defined as the extension of first-order logic by quantification over sets of edges and sets of vertices. That is, in addition to first-order variables there are variables $X, Y, \ldots$ ranging over sets of vertices and variables $F, F', \ldots$ ranging over sets of edges. Formulas of MSO$_2[\Sigma]$ are then build up inductively by the rules for first-order logic with the following additional rules: if $X$ is a second-order variable either ranging over a set of vertices or a set of edges and $\varphi \in \mathrm{MSO}_2[\Sigma \dot\cup \{X\}]$, then $\exists X \varphi \in \mathrm{MSO}_2[\Sigma]$ and $\forall X \varphi \in \mathrm{MSO}_2[\Sigma]$ where, e.g., a formula $\exists F \varphi$, $F$ being a variable over sets of edges, is true in a $\Sigma$-graph $G$ if there is a subset $F' \subseteq E(G)$ such that $\varphi$ is true in $G$ if the variable $F$ is interpreted by $F'$. We write $G \models \psi$ to indicate that a formula $\psi$ is true in $G$. See [Lib04] for more on MSO$_2$.

We are primarily interested in the complexity of checking a fixed formula expressing a graph property in a given input graph. We therefore study model-checking problems in the framework of *parameterized complexity* (see [FG06] for background on parameterized complexity). Let $\mathcal{C}$ be a class of $\Sigma$-graphs. The *parameterized model-checking problem* MC(MSO$_2, \mathcal{C}$) for MSO$_2$ on $\mathcal{C}$ is defined as the problem to decide, given $G \in \mathcal{C}$ and $\varphi \in \mathrm{MSO}_2[\sigma]$, if $G \models \varphi$. The *parameter* is $|\varphi|$. MC(MSO$_2, \mathcal{C}$) is *fixed-parameter tractable* (fpt), if for all $G \in \mathcal{C}$ and $\varphi \in \mathrm{MSO}_2[\sigma]$, $G \models \varphi$ can be decided in time $f(|\varphi|) \cdot |G|^k$, for some computable function $f$ and $k \in \mathbb{N}$. The problem is in the class XP, if it can be decided in time $|G|^{f(|\varphi|)}$. As, for instance, the NP-complete problem 3-Colourability is definable in MSO$_2$, MC(MSO$_2$, GRAPHS), the model-checking problem for MSO$_2$ on the class of all graphs, is not fixed-parameter tractable unless $P = \mathrm{NP}$. However, Courcelle proved that if we restrict the class of admissible input graphs, then we can obtain much better results.

**Theorem 6.1** ([Cou90]). MC(MSO$_2, \mathcal{C}$) *is fixed-parameter tractable on any class $\mathcal{C}$ of graphs of treewidth bounded by a constant.*

Courcelle's theorem gives a sufficient condition for MC(MSO$_2, \mathcal{C}$) to be tractable. We now show that on coloured graphs, Courcelle's theorem can not be extended much further. We first need some definitions.

The treewidth of a class $\mathcal{C}$ of graphs is *strongly unbounded* by a function $f : \mathbb{N} \to \mathbb{N}$ if there is a polynomial $p(x)$ such that for all $n \in \mathbb{N}$

1. there is a graph $G_n \in \mathcal{C}$ of treewidth between $n$ and $p(n)$ whose treewidth is not bounded by $f(|G_n|)$
2. given $n$, $G_n$ can be constructed in time $2^{n^\varepsilon}$, for some $\varepsilon < 1$.

The treewidth of $\mathcal{C}$ is *strongly unbounded poly-logarithmically* if it is strongly unbounded by $\log^c n$, for all $c \geq 1$. Essentially, *strongly* means that a) there are not too big gaps between the treewidth of graphs witnessing that the treewidth of $\mathcal{C}$ is not bounded by $f(n)$ and b) we can compute such witnesses efficiently. This is needed because the proof of the theorem below relies on a reduction of an NP-complete problem $P$ to MC(MSO$_2, \mathcal{C}$) so that given a word $w$ for which we want to decide if $w \in P$ we construct a graph $G_w$ of treewidth polynomial in $|w|$ and whose treewidth is $> \log^{24} |G|$. If $\mathcal{C}$ was not strongly unbounded then there simply would not be enough graphs of large treewidth in $\mathcal{C}$ to define any reduction.

The following theorem was proved in [Kre09]. Let $\Gamma$ be a set of colours with at least one edge and two vertex colours.

**Theorem 6.2** ([Kre09]). *Let $\mathcal{C}$ be a constructible class of $\Gamma$-coloured graphs closed under colourings.*

1. *If the treewidth of $\mathcal{C}$ is strongly unbounded poly-logarithmically then $\mathrm{MC}(\mathrm{MSO}_2, \mathcal{C})$ is not in XP, and hence not fpt, unless all problems in NP (in fact, all problems in the polynomial-time hierarchy) can be solved in sub-exponential time.*

2. *If the treewidth of $\mathcal{C}$ is strongly unbounded by $\log^{16} n$ then $\mathrm{MC}(\mathrm{MSO}_2, \mathcal{C})$ is not in XP unless SAT can be solved in sub-exponential time.*

Here, a class $\mathcal{C}$ is called *constructible* if given a graph $G \in \mathcal{C}$ of treewidth $c \cdot l^8 \cdot \sqrt{\log(l^2)}$, for some constant $c$ defined in [Kre09], we can compute in polynomial time a structure called a *coloured pseudo-wall of order $m$*. A coloured pseudo-wall of order $m$ is a variant of a grid-like minor and can easily be computed from a given grid-like minor of order $m$. Using Theorem 4.3, we can now compute grid-like minors and hence pseudo-walls in plolynomial time, at the expense that the graph in which we compute these structures needs to have treewidth at least $c'_2 l^{12}$ instead of $c \cdot l^8 \cdot \sqrt{\log(l^2)}$. Hence, we obtain the following result.

**Theorem 6.3.** *Let $\mathcal{C}$ be any class of $\Gamma$-coloured graphs closed under colourings.*

1. *If the treewidth of $\mathcal{C}$ is strongly unbounded poly-logarithmically then $\mathrm{MC}(\mathrm{MSO}_2, \mathcal{C})$ is not in XP, and hence not fpt, unless all problems in NP (in fact, all problems in the polynomial-time hierarchy) can be solved in sub-exponential time.*

2. *If the treewidth of $\mathcal{C}$ is strongly unbounded by $\log^{24} n$ then $\mathrm{MC}(\mathrm{MSO}_2, \mathcal{C})$ is not in XP unless SAT can be solved in sub-exponential time.*